\documentstyle[preprint,aps]{revtex}
\font\elevenbf=cmbx10 scaled\magstep 1
\newcommand{\shat}{\mbox{$\hat{s}$}}

\newcommand{\rs}{\mbox{$\sqrt{s}$}}
\newcommand{\gaga}{\mbox{$\gamma\gamma$}}
\newcommand{\be}{\begin{equation}}
\newcommand{\ene}{\end{equation}}
\newcommand{\een}{\end{subequations}}
\newcommand{\ben}{\begin{subequations}}
\newcommand{\bea}{\begin{eqnarray}}
\newcommand{\eea}{\end{eqnarray}}
\newcommand{\MRSD}{\rm MRSD}
\begin{document}
\preprint{\vbox{\baselineskip=14pt
  \rightline{MAD/PH/824} \break
  \rightline{PRL-TH-94/10} \break
  \rightline{BU-TH-94/3} \break
  \rightline{March 1994} }}
\title
{\bf \gaga\ PROCESSES AT HIGH ENERGY pp COLLIDERS\\}
\author{Manuel Drees\footnote{Heisenberg Fellow}}
\address{ Physics Department, University of Wisconsin, Madison,
WI 53706, USA\\}
\author{Rohini M. Godbole}
\address {Physics Department, University of Bombay, Vidyanagari,
Bombay 400098, India\\}
\author {Marek Nowakowski\footnote{Feodor-Lynen Fellow}  and
Saurabh D. Rindani}
\address{Theory Group, Physical Research Laboratory, Navrangpura,
Ahmedabad 380 009,India\\}
\maketitle
\begin{abstract}

In this note we investigate the production of charged heavy particles via
\gaga\ fusion at high energy pp colliders. We revise previous claims that the
\gaga\ cross section is comparable to or larger than that for the
corresponding Drell-Yan process at high energies. Indeed we find that the
\gaga\ contribution to the total production cross section at pp is far below
the Drell-Yan cross section. As far as  the individual elastic, semi-elastic
and inelastic contributions to the \gaga\ process are concerned we find that
they are all of the same order of magnitude.
\end{abstract}
\newpage

The detection of a fundamental charged scalar particle would certainly lead
beyond the realm of the Standard Model (SM). These
particles can arise either in the context of supersymmetric
models, as superpartners of quarks and leptons \cite{susy}, or in
extended Higgs sectors, e.g. in two-Higgs-doublet models
\cite{higgs2} (with or without supersymmetry) or in models with triplet
Higgses \cite{triplet}. In general, the different charged
scalars will have different interactions at tree level. For
instance, sleptons do not couple to quarks in contrast to
$H^{\pm}$ in the two Higgs doublet model, while one charged Higgs boson
in triplet models does not couple to matter at all but has an
unconventional $H^+ W^- Z^0$ vertex. Hence a model independent
production mechanism is welcome. Such a model independent
interaction is clearly given by the scalar QED part of the
underlying theory. For example the \gaga\ fusion
processes:
\be \label{e1}
\gaga\ \to H^+ H^-, \;\; \tilde{l}^+ \tilde{l}^-, \;\;... \ene
are uniquely calculable for given mass of the produced particles.
At pp colliders we also have, however, the possibility of the
$q\overline{q}$ annihilation Drell--Yan processes
\be \label{e2}
q\overline{q} \to H^+ H^-, \; \; \tilde{l}^+ \tilde{l}^-\;\; ... \ene

There has been a claim in the literature that the \gaga\ fusion exceeds the
Drell--Yan (DY) cross sections at pp by orders of magnitude \cite{choi}. This
would be an interesting possibility of producing charged heavy scalars at
hadronic colliders or for that matter any charged particle which does not have
strong interactions.

Apart from the charged scalars mentioned above there exist various candidates
for  charged fermions.  These fermions can be either fourth generation
leptons, charginos or exotic leptons in extended gauge theories like $E_6$
\cite{exotics}. Current  limits on the masses of all exotic charged particles
which couple to the Z with full strength are $\sim M_Z/2$. In the case of
$H^{\pm}$ there exist additional constraints (clearly model dependent) from
the experimental studies of the $b \to s \gamma $ decay. In one variation of
the model $m_{H^{\pm}} < 110$ GeV is ruled out for large values of $\tan
\beta$ and for $m_t = 150$ GeV \cite{btogamma}.  However, in the
two-Higgs-doublet models with SUSY these constraints are much
weaker\cite{guidice}. (The same analysis also shows that there are no limits
on the chargino masses from the $b \to s \gamma $ rate.) The calculation for
$\gaga\ \to L^+ L^-$ at pp colliders has been done recently \cite{kaly}. The
result in \cite{kaly} is that the \gaga\ cross section is comparable to the
corresponding Drell--Yan process at high energies, e.g. at $ \rs\ = 40$ TeV for
$m_L \sim 100$ GeV. At LHC energies the \gaga\ cross section in the same mass
range was found to be \cite{kaly} one order of magnitude smaller than the DY
cross section.

We have repeated the calculations for scalar and fermion pair production, and
find that in both cases the \gaga\ cross sections are well below the
Drell--Yan contribution \cite{note2}. In what follows we outline briefly the
basic tools and approximations in the calculation.

In order to calculate the pp cross section we have used the
Weizs\"acker-Williams approximation \cite{equival} for the inelastic case
($\gamma p X$ vertex)  and a modified version of this approximation
\cite{kniehl,manuel} for the elastic case  ($\gamma p p$ vertex). In the
latter case the proton remains intact. The inelastic (inel.) total pp cross
section for $H^+ H^-$ as well as $L^+ L^-$ production reads
\bea \label{e3}
\sigma_{pp}^{inel.}(s)&=& \sum_{q,\; q'} \int_{4m^2/s}^1dx_1
\int_{4m^2/sx_1}^1 dx_2 \int_{4m^2/sx_1 x_2}^1 dz_1
\int_{4m^2/sx_1 x_2 z_1}^1 dz_2 \;\;e_q^2 e_{q'}^2 \nonumber \\
& \cdot &f_{q/p}(x_1,\;Q^2)\;f_{q'/p}(x_2,\; Q^2)f_{\gamma
/q}(z_1)\;f_{\gamma /q'}(z_2) \;\hat{\sigma}_{\gamma \gamma}(x_1 x_2 z_1 z_2s)
\eea
where $m$ is the mass of either $H^{\pm}$ or $L^{\pm}$, $e_u=2/3$,
$e_d=- 1/3$ and $\hat{\sigma}_{\gamma \gamma}$ is the production
subprocess cross section with the center of mass energy
$\sqrt{\hat{s}} = \sqrt{x_1 x_2 z_1 z_2s}$.  The structure functions
have the usual meaning: $f_{q/p}$ is the quark density  inside
the proton, $f_{\gamma / q}$ is the photon spectrum inside a
quark. We use the the $\MRSD _{-}^{'}$ parameterization for the partonic
densities inside the proton \cite{roberts}. The scale $Q^2$ has
been chosen throughout the paper to be $\hat{s}/4$. With
\be \label{e4}
f_{\gamma}(z) \equiv f_{\gamma /q}(z)=f_{\gamma /q'}(z)
={\alpha_{em} \over 2 \pi}\;\;{(1+(1-z)^2) \over z}\;\;\ln(Q_1^2/Q_2^2)
\ene
we can write (\ref{e3}) in a more compact form as
\bea \label{e5}
\sigma_{pp}^{inel.}(s)&=&\int_{4m^2/s}^1dx_1 \int_{4m^2/sx_1}^1dx_2
\int_{4m^2/sx_1 x_2}^1dz_1 \int_{4m^2/sx_1 x_2 z_1}^1dz_2\;\; {1
\over x_1}F_2^p(x_1,\;Q^2) \nonumber \\
&\cdot &{1 \over x_2}F_2^p(x_2,\;Q^2)
f_{\gamma}(z_1)\;f_{\gamma}(z_2)\;\hat{\sigma}_{\gamma
\gamma}(x_1 x_2 z_1 z_2 s) \eea
where $F_2^p$ is the deep-inelastic proton structure function.
There is a certain ambiguity about the choice of the scales $Q_i^2$ in
the argument of the log in eq. (\ref{e4}). We choose $Q_1^2$ to be the
maximum value of the momentum transfer given by $\shat/4 - m^2$ and
the choice of $Q_2^2 = 1$ GeV$^2$ is made such that the photons are
sufficiently off--shell for the Quark--Parton--Model to be applicable.

The semi-elastic (semi-el.) cross section for $pp \to H^+ H^-
(L^+ L^-)pX$ is given by
\bea \label{e7}
\sigma_{pp}^{semi-el.}(s)&=&2\int_{4m^2/s}^1dx_1\int_{4m^2/sx_1}^1dz_1
\int_{sm^2 /sx_1 z_1}^1dz_2 {1 \over
x_1}\;\;F_2^p(x_1,\;Q^2) \nonumber \\
& \cdot & f_{\gamma}(z_1)
\;f_{\gamma /p}^{el.}(z_2)\;\hat{\sigma}_{\gamma \gamma}(x_1 z_1
z_2 s) \eea
The subprocess energy now is given by $\sqrt{\hat s} = \sqrt{sx_1 z_1z_2} $.
The elastic photon spectrum $f_{\gamma /p}^{el.}(z)$ has been obtained in the
form of an integral in \cite{kniehl}. However, we use an approximate analytic
expression given in \cite{manuel} which is known to reproduce exact results to
about 10\%. The form we use is given by
\be \label{e8}
f_{\gamma /p}^{el.}(z)={\alpha_{em} \over 2 \pi z}(1
+(1-z)^2)\left[\ln A -{11 \over 6} +{3\over A}-{3 \over 2A^2}+{1
\over 3A^3}\right] \ene
where
\be \label{e9}
A=1+{0.71({\rm GeV})^2 \over Q_{min}^2} \ene
with
\bea \label{e10}
Q_{min}^2&=&-2m_p^2 + {1 \over 2s}\biggl[(s+m_p^2)(s-zs+m_p^2)
\nonumber \\
&-& (s-m_p^2)
\sqrt{(s-zs-m_p^2)^2- 4m_p^2zs}\biggr] \eea
At high energies $Q_{min}^2$ is given to a very good approximation
by $m_p^2 z^2/(1-z)$. Since the relevant values of the scaled
photon energy $z_i$ can in general take smaller values in
the elastic  case as compared to the inelastic case, eqs.
(\ref{e10}),(\ref{e9}) and (\ref{e8}) imply that even in the
elastic case there is a logarithmic enhancement of the photon densities.

Finally the pure elastic contribution wherein both the photons
remain intact and hence can in principle give rise to clean
events, can  be written as
\be \label{e11}
\sigma_{pp}^{el.}(s)=\int_{4m^2/s}^1 dz_1 \int_{4m^2/z_1s}^1
dz_2 \;\;f_{\gamma /p}^{el.}(z_1)\;f_{\gamma /p}^{el.}(z_2)\;
\hat{\sigma}_{\gamma \gamma}(\hat s = z_1z_2s) \ene

Defining $\hat{\beta}_{L,\;H}=(1 -4m^2_{L,\;H}/\hat s)^{1/2}$ the \gaga\
subcross sections take the simple form
\be \label{e12}
\hat{\sigma}(\gamma \gamma \to H^+ H^-) ={2 \pi
\alpha_{em}^2 (M^2_W)  \over \hat{s}} \hat{\beta}_H\left[2 - \hat{\beta}^2_H
-{1 -\hat{\beta}_H^4 \over 2 \hat{\beta}_H} \ln {1 + \hat{\beta}_H
\over 1- \hat{\beta}_H}\right], \ene
and for lepton production
\be \label{e13}
\hat{\sigma}(\gamma \gamma \to L^+ L^-)={4 \pi
\alpha_{em}^2(M_W^2) \over \hat{s}}\hat{\beta}_L\left[{3-\hat{\beta}_L^4
\over 2\hat{\beta}_L} \ln {1 + \hat{\beta}_L \over 1 -\hat{\beta}_L}
-(2-\hat{\beta}_L^2)\right]. \ene
 Note that we have used  $\alpha_{em}=1/137$ in
(\ref{e4}) and (\ref{e8}) and $\alpha_{em}(M_W^2)=1/128$ in the
subcross sections (\ref{e12}) and (\ref{e13}).

For completeness we also give here the Drell-Yan $q\overline{q}$
annihilation cross section to $H^+H^-$ including $Z^0$ exchange, for the
case that $H^{\pm}$ resides in an $SU(2)$ doublet:
\bea \label{e14}
\hat{\sigma}(q\overline{q} \to H^+ H^-)&=& {4 \pi \alpha_{em}^2(M_W^2)
\over
3\hat{s}}\;\; {(\hat{\beta}_H)^{3/2} \over 4}\biggl[e^2_q + 2e_q g_{V_q}
{\cot 2\theta_W \over \sin 2\theta_W} \;\; {\hat{s}(\hat{s}-m_Z^2)
\over (\hat{s}-m_Z^2)^2
+ \Gamma^2_Z m_Z^2} \nonumber \\
&+&(g_{V_q}^2 + g_{A_q}^2) {\cot ^2 2\theta_W \over \sin^2 2\theta_W}
\;\; {\hat{s}^2 \over (\hat{s}-m_Z^2)^2 +\Gamma^2_Z m_Z^2 }\biggr] \eea
In the above $g_{V_q}$ ,  $g_{A_q}$ are the standard vector and
axial vector coupling for the quark.

The results of our calculations are presented in Fig. \ref{fig.1} for $H^+
H^-$ production and in Fig. \ref{fig.2} for the lepton case. As far as the
$H^+ H^-$ production in \gaga\ fusion is concerned we differ from the results
given in \cite{choi} by roughly three orders of magnitude: our \gaga\ cross
section is far below their results and also approximately two orders of
magnitude smaller than the DY cross section. The logarithmic enhancement of
the photon densities is simply not enough to overcome completely the extra
factor $\alpha_{em}^2$ in the \gaga\ process. Even if the Higgs is doubly
charged (such a Higgs appears in triplet models \cite{triplet}) the ratio of
DY to \gaga\ cross section changes only by a factor $1/4$ as compared to the
singly charged Higgs production. We also find that contributions from elastic,
semi-elastic and inelastic processes to the total \gaga\ cross section are of
the same order of magnitude. The elastic process contributes $\sim 20\% $ of
the total \gaga\ cross--section at smaller values of $m_H$ going up to $30\%$
at the high end. This can be traced to the logarithmic enhancement of the
photon density even in the  elastic case mentioned earlier. Assuming the
$\tilde l_L, \tilde l_R$ to be degenerate in mass the cross--section for
\gaga\  production  of sleptons (for one generation) will be twice the
corresponding $H^+H^-$ cross--sections.

Our results for leptons are given in Fig.\ref{fig.2}. Here again we find that
at LHC energies DY exceeds \gaga\ by two orders of magnitude even for
relatively small $m_L$ masses in the range of $50-100$ GeV\cite{note2}. In
general the $ L^+L^-$ cross--sections are higher than the corresponding
$H^+H^-$ cross--sections (both for \gaga\ and DY) by about a factor of 5--7.
This can be traced to the different spin factors and the different  $\hat
\beta$ dependence of the subprocess cross--section for the fermions and
scalars.  The cross--section for the \gaga\ production of charginos  will
again be the same as that of the charged leptons.

One might think that by sacrificing rate for `cleanliness' the
purely elastic processes might prove useful. Moreover, even inelastic or
semi--elastic \gaga\ events might be characterized by ``rapidity gaps", where
the only hadrons at central rapidities are due to the decay of the heavy
particles produced. However, at the LHC one expects about 16 minimum bias
events per bunch crossing at luminosity ${\cal L} = 10^{34}$ cm$^{-2}$
sec$^{-1}$; even the elastic \gaga\ events will therefore not
be free of hadronic debris. These ``overlapping events" will fill the entire
rapidity space with (mostly soft) hadrons, thereby obscuring any rapidity gap.
Notice also that in the purely elastic events the participating protons only
lose about 0.1\% of their energy, making it very difficult to detect them in
a forward spectrometer of the type now being installed at HERA detectors.
We are therefore forced to conclude that most likely one will not be able
to distinguish experimentally between DY and \gaga\ events if the LHC is
operated anywhere near its design luminosity. The clean elastic events might
be detectable at luminosities well below 10$^{33}$ cm$^{-2}$ sec$^{-1}$,
where event overlap is not expected to occur. However, our results show that
at such a low luminosity one is running out of event rate at masses not much
above the limit that can be probed at LEP200; moreover, there might be
sizable backgrounds, e.g. due to the process $\gaga \rightarrow W^+ W^-$.

At this point it might be instructive to compare the \gaga\ cross--sections
with other (model dependent) possible production mechanisms for various weakly
interacting charged particles. Studies \cite{barbier1} have shown that search
for charginos in hadronically quiet multilepton events due to associated
production of a chargino with a neutralino (via DY) at LHC might be feasible
up to $m_{\chi^\pm} \simeq 250$ GeV. The detection of sleptons with mass up to
$\sim 250$ GeV also seems possible \cite{slep}. Hence the DY process still
seems to be the dominant mode for production for sleptons, charginos as well
as heavy leptons. For larger masses the DY cross--section falls off and in
some cases the gluon induced production (which we discuss below) will take
over.

For the charged Higgses the situation is somewhat different. The question of
DY/\gaga/gg production becomes relevant in this case only for $m_t <
m_{H^\pm}$. If $m_t > m_{H^\pm}$ the charged Higgs can be produced in the
decay of the top quark and the strong production of top quarks gives large
rates allowing one to probe at LHC upto $m_{H^\pm} \sim m_t - 20$ GeV
\cite{dp1}. Even when $m_t < m_{H^\pm}$ production of a single charged
Higgs in association with a $t$ quark via the process
\bea \label{e16}
gb \to t H^- \eea
might provide a measurable signal in the decay channel
\be \label{e17}
t H^+ \to ttb \to b (b q \bar q^{\prime}) (b l \nu).
\ene
The cross--section is $\sim 15 $ pb for $m_{H^\pm} \sim 150$ GeV and could
provide a feasible signal up to $m_{H^\pm} \sim 200$ GeV over a wide range of
parameter space, if $b$ quarks can be tagged with high purity and not too low
efficiency \cite{dp}. Fig.\ref{fig.1} shows that even for the DY process the
charged Higgs cross--section is only a few tens of fb or less if $m_{H^+} >
m_t$.

Another process that contributes to the pair production of Higgs bosons and
charged leptons is one loop gluon fusion:
\bea \label{e15}
&gg& \to H^+ H^- \nonumber \\
&gg& \to L^+ L^- \eea
These contributions will only be competitive with ordinary DY production if
some couplings of the produced particles grow with their mass. Accordingly the
first process will be large \cite{willen} if $m_t > m_{H^+}$ (in which case
$H^+$ production from $t$ decays will have even larger rates) but is expected
to decrease for $m_{H^+} > m_t$. Since the coupling of chiral leptons to Higgs
bosons and longitudinal Z bosons grows with the lepton mass, graphs containing
the (1--loop) $ggH^{0*}$ and $ggZ^{0*}$ couplings dominate the production of
both charged \cite{dicus} and neutral \cite{apost} chiral leptons of
sufficiently large mass.

In summary, we have shown that the cross section for the pair production of
heavy charged scalars or fermions via \gaga\ fusion amounts to at best a few
\% of the corresponding Drell--Yan cross section; in many cases there are
additional production mechanisms with even larger cross sections. Moreover, at
the LHC overlapping events prevent one from isolating \gaga\ events
experimentally unless the machine is run at a very low luminosity, in which
case the accessible mass window is not much larger than at LEP200. We do
therefore not expect \gaga\ fusion processes at the LHC to be competitive with
more traditional mechanisms for the production of new particles.

While writing this note we have received a preprint \cite{zerwas} which treats
the same subject of \gaga\ processes in pp colliders and gets similar results.
However we differ somewhat in the details which is most probably due to the
different treatment of the photon luminosity functions.

{\elevenbf \noindent Acknowledgments \hfil}
\vglue 0.4cm
We thank the Department of Science and Technology (India)  and the organizers
of the WHEPP-III workshop held in Madras (India) where this work was partially
done. The work of M.D. was supported by a grant from the Deutsche
Forschungsgemeinschaft under the Heisenberg program, as well as in part by the
U.S. Department of Energy under contract No. DE-AC02-76ER00881, by the
Wisconsin Research Committee with funds granted by the Wisconsin Alumni
Research Foundation, and by the Texas National Research Laboratory Commission
under grant RGFY93--221. R.M.G. wishes to thank the Theory Group at PRL for
hospitality. M.N. wishes to thank the Alexander von Humboldt foundation for
financial support under the Feodor-Lynen fellowship program.

\begin{figure}
\caption{Cross section in fb for DY and \gaga\ production of the charged
Higgs at LHC energies, as a function of the Higgs mass. The dashed,
dash-dotted and long-dashed lines show the el., inel. and semi-el.
contributions (as defined in the text) to the \gaga\ cross sections.
The total \gaga\ cross section and the DY contributions are shown
by the labeled solid lines.}
\label{fig.1}
\end{figure}
\begin{figure}
\caption{Cross section in fb for DY and \gaga\ production of the charged
Leptons at LHC energies, as a function of the Lepton mass. The
convention is the same as in \protect\ref{fig.1}.}
\label{fig.2}
\end{figure}
\end{document}